\documentclass[%
    reprint,
    superscriptaddress,
    longbibliography,
    nofootinbib,
    amsmath,
    amssymb,
    aps,
    floatfix,
]{revtex4-2}

\usepackage[T1]{fontenc}
\usepackage[utf8]{inputenc}
\usepackage{amsmath, amsfonts, amssymb, mathtools, braket, bm}
\usepackage{graphicx}
\usepackage{tikz}
\usetikzlibrary{quantikz2}
\usepackage{booktabs}
\usepackage{multirow}
\usepackage{dcolumn}
\usepackage{enumitem}
\usepackage{nicefrac}
\usepackage{microtype}
\usepackage{xspace}
\usepackage{listings}
\usepackage{siunitx}
\usepackage{acronym}
\usepackage[hidelinks]{hyperref}

\NewDocumentCommand{\codeword}{v}{%
\texttt{\textcolor{blue}{#1}}%
}

\setlist[itemize]{label=-}

\begin{document}

\title[Tensor Network for Anomaly Detection in the Latent Space]{%
Tensor Network for Anomaly Detection\\in the Latent Space of Proton Collision Events at the LHC}
% \title{Enhancing Machine Learning Applications through Optimized Matrix Product Structures}
\author{Ema Puljak}\thanks{Contact author: \href{mailto:ema.puljak@cern.ch}{ema.puljak@cern.ch}}
\affiliation{Departamento de Física, Universitat Autònoma de Barcelona, 08193 Bellaterra (Barcelona), Spain}
\affiliation{Barcelona Supercomputing Center, 08034 Barcelona, Spain}
\author{Maurizio Pierini}
\affiliation{European Organization for Nuclear Research (CERN), CH-1211 Geneva, Switzerland}
\author{Artur Garcia-Saez}
\affiliation{Barcelona Supercomputing Center, 08034 Barcelona, Spain}
\affiliation{Qilimanjaro Quantum Tech, 08019 Barcelona, Spain}

\begin{abstract}
The pursuit of discovering new phenomena at the Large Hadron Collider (LHC) demands constant innovation in algorithms and technologies. Tensor networks are mathematical models on the intersection of classical and quantum machine learning, which present a promising and efficient alternative for tackling these challenges. In this work, we propose a tensor network-based strategy for anomaly detection at the LHC and demonstrate its superior performance in identifying new phenomena compared to established quantum methods. Our model is a parametrized Matrix Product State with an isometric feature map, processing a latent representation of simulated LHC data generated by an autoencoder. Our results highlight the potential of tensor networks to enhance new-physics discovery.
\end{abstract}

\maketitle

\section{Introduction}
Tensor Network (TN) models, originating from condensed matter physics~\cite{MPS-PEPS-VRGM, ORUS2014117, TN_book} to describe highly-correlated quantum systems, have been recently explored as well in the context of high-energy physics (HEP)\cite{Dalmonte02072016, PhysRevB.96.195123, sim-lattice20, Carmen_Banuls_2020, Montangero_2021, q-inspired-hep21, meurice2022tensornetworkshighenergy, magnifico2024tensornetworkslatticegauge}. However, their applications in HEP have primarily focused on quantum information processing, lattice field theory, and b-jet tagging. Given their ability to efficiently capture and analyze complex relationships in high-dimensional data and their proven success in solving machine learning (ML) problems~\cite{stoudenmire2017supervised, Reyes_2021, Ran_2023, tomut2024compactifaiextremecompressionlarge}, TNs present a promising direction for ML applications in HEP. Recent studies have investigated their use in HEP for classification ~\cite{q-inspired-hep21} and event reconstruction~\cite{Araz_2021}.

To explore new physics phenomena at the Large Hadron Collider (LHC), proton collisions are analyzed to identify deviations from events predicted by the Standard Model (SM) of particle physics. Anomaly detection, as an ML task, plays an important role as a signal-agnostic approach searching for anomalous or unforeseen signatures beyond those described by the SM~\cite{Canelli_2022, Govorkova_2022, gandrakota2024realtimeanomalydetectionl1, BELIS2024100091}. In recent years, the ATLAS collaboration at CERN has introduced models based on weak-supervised learning~\cite{Aad_2020} and anomaly detection~\cite{Aad_2023, Aad_2024} strategy, while the CMS experiment has integrated deep learning anomaly detection techniques directly into its data selection pipelines~\cite{cmscollaboration2024modelagnosticsearchdijetresonances}. A common ML-based application in HEP is the search for new physics in the di-jet final state, explored in recent studies~\cite{Collins_2018, Farina_2020, Kasieczka_2021, Quak, PhysRevD.105.115009, golling2023massiveissueanomalydetection}. In high-energy collisions, jets are formed when quarks and gluons hadronize into detectable particles. These particles, measured as clustered energy deposits and tracks in specific detector components, allow the jet to be reconstructed using algorithms to group them based on their distance in angular space. One-prong, cone-shaped jets are abundantly produced in Quantum Chromodynamics (QCD) multijet events, representing the highest-rate phenomena produced in the LHC collisions (background). In contrast, multi-prong jets may originate from the decay of heavy particles into multiple quarks and/or gluons, leading to a resonance peak in the dijet-mass ($m_{jj}$) spectrum. Detecting such peaks is challenging due to a substantial multijet background. Anomaly detection methods can be employed to enhance the signal-to-background ratio and improve sensitivity to beyond-the-Standard-Model (BSM) signals. These techniques are designed to identify statistically rare features in the data that deviate from the expected background. At the LHC, efficient data selection and filtering are achieved through a two-stage trigger system. The first stage, the Level-1 Trigger (L1T), is fully based on low-latency hardware, such as Field Programmable Gate Arrays (FPGAs), running selection algorithms within $\mathcal{O}(1)\mu \text{s}$. The second stage, the High Level Trigger (HLT), is a software-based system implemented in the computing farm with a latency requirement of $\mathcal{O}(100)\text{ms}$~\cite{BELIS2024100091}. To enable efficient online deployment and make this task more tractable, detecting new-physics signatures in the latent space of the autoencoder model has been proposed as a realistic strategy for the future LHC trigger system~\cite{ PhysRevD.107.016002, VAE-LHC, Govorkova_2022, Belis2024}. This approach creates compact representations by reducing the dimensionality of the problem while preserving important features of the collision events, where anomaly detection can be more effective. In this work, as input representation, we consider the latent space of di-jet events, as described in Ref.~\cite{Belis2024}.

\begin{figure*}[htbp]
    \centering
    \includegraphics[width=\textwidth]{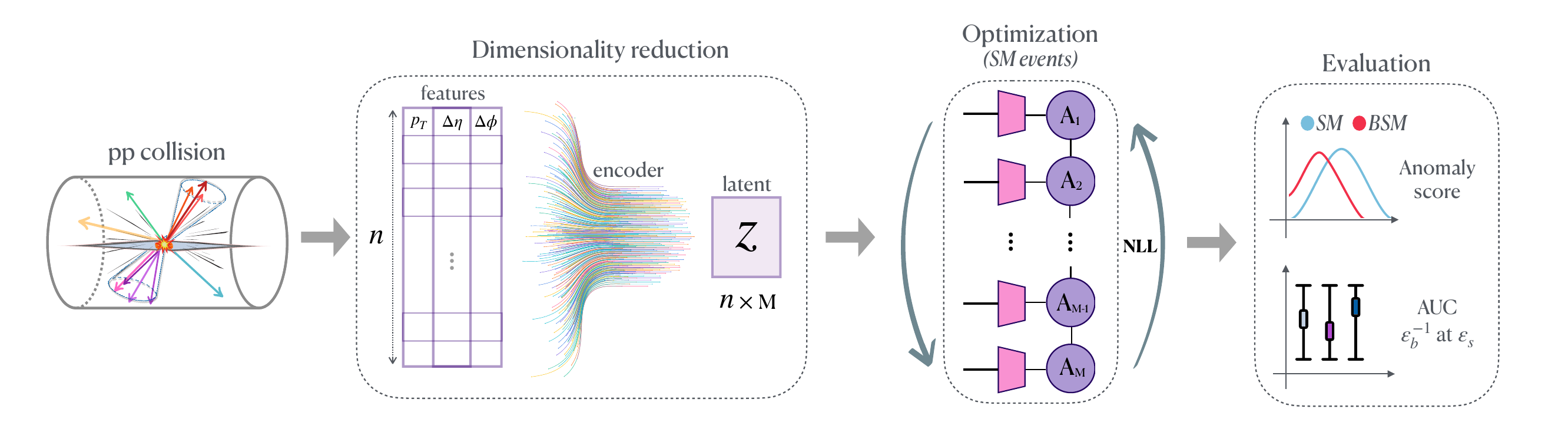}
    \caption{\textbf{Anomaly detection pipeline}: Proton-proton collision data ($n$ samples with $p_T, \Delta\eta, \Delta\phi$ features) is passed through an encoder, generating $M$ latent features mapped by isometric feature functions into a product state contracted with an MPS model with $A_k$ trainable parameters (purple circle). The MPS, trained on SM samples via NLL loss minimization, is evaluated both on SM and BSM events. Performance is measured using output probabilities as anomaly scores and metrics from the Receiver Operating Characteristic (ROC) curve (Area Under the Curve (AUC) value and background efficiency, $\varepsilon_b$, at a given signal efficiency, $\varepsilon_s$).}
    \label{fig:pipeline}
\end{figure*}

Collision event data from the LHC originates from quantum processes, which have recently sparked significant interest in quantum machine learning (QML) approaches for HEP\cite{Mott2017, caldeira2020restrictedboltzmannmachinesgalaxy, T_ys_z_2020, Terashi2021, Chang_2021, Guan_2021, alvi_quantum_2022, PhysRevD.106.036021, PhysRevD.106.096006, Schenk_2024}. Among these, quantum anomaly detection approaches have been proposed to search for new particles\cite{Belis2024, alvi_quantum_2022, PhysRevD.105.095004}, following an emerging trend in LHC-related literature. However, the practical deployment of quantum algorithms remains challenging on current hardware, making the direct implementation at the LHC unrealistic for the time being. As an alternative approach, one can use quantum-inspired Tensor Networks, leveraging the quantum characteristics of the data and offering the possibility of future deployment within the collider system at CERN~\cite{borella2024ultralowlatencyquantuminspiredmachine}.

In this work, we present a quantum-inspired Tensor Network model for a realistic use case of anomaly detection in the latent space of proton collision events at the LHC. Specifically, we employ a one-dimensional parameterized Matrix Product State (MPS)~\cite{MPS-PEPS-VRGM, MPS_cite} to learn a probability distribution over continuous latent variables produced by the autoencoder, identifying deviations from this learned distribution as anomalies (see Fig.~\ref{fig:pipeline}). This particular anomaly detection problem was previously explored through a systematic study of quantum anomaly detection methods in Ref.~\cite{Belis2024}, which we use as a benchmark to validate our TN-based approach. The extension of TN models for probabilistic modeling in the continuous regime was introduced in Ref.~\cite{meiburg2024generativelearningcontinuousdata} and studied for different data domains in Ref.~\cite{hohenfeld2025explaininganomaliestensornetworks}. Furthermore, we assess the runtime performance of the complete anomaly detection pipeline consisting of the encoder and the MPS model, to evaluate its suitability for online deployment at the LHC. Our results highlight the efficiency and effectiveness of TNs for this task, supporting their potential integration as quantum-inspired ML models for real-time anomaly detection in the LHC trigger system.

This paper is organized as follows: Sec.~\ref{sec:methodology} introduces the dataset and dimensionality reduction technique, followed by a description of the theoretical foundation of TNs and their application to discrete probabilistic modeling. Next, we present our TN model for anomaly detection along with the training methodology. Sec.~\ref{sec:results} provides a detailed systematic analysis of our experiments, and Sec.~\ref{sec:conclusion} emphasizes the significance of our approach as a viable quantum-inspired model with potential for deployment in the real-time selection data flow at the LHC.

\section{Methodology}\label{sec:methodology}
\subsection*{Dataset}
This study is based on a dataset of simulated dijet events, generated with $\mathrm{PYTHIA}$~\cite{SJOSTRAND2015159} library, at fixed center-of-mass energy of $\sqrt{\mathrm{s}} = 13 \: \mathrm{TeV}$. The $\mathrm{PYTHIA}$ events are passed to a $\mathrm{DELPHES}$-based simulation~\cite{DELPHES3} of the CMS events, to emulate CMS detector resolution and efficiency effects, and are further processed by a $\mathrm{DELPHES}$ implementation of the CMS particle-flow reconstruction~\cite{Sirunyan_2017}. The training dataset is defined using a dijet pseudorapidity separation method, specifically focusing on events with a jet-to-jet separation value of $|\Delta\eta_{jj}| > 1.4$ between the two jets with the highest $p_T$ value. This configuration primarily captures QCD multijet events with minimal expected contamination from potential BSM collision processes. Background or SM events are generated by simulating QCD multijet production, the most dominant set of events at the LHC.  The new-physics BSM processes used in this study are considered as potential anomalies:
\begin{itemize}
    \item Production of narrow Randall-Sundrum gravitons decaying to two W-bosons~\cite{Randall_1999}($\mathrm{NA \ G \rightarrow WW}$).
    \item Production of broad Randall-Sundrum gravitons decaying to two W-bosons~\cite{Randall_1999}($\mathrm{BR \ G \rightarrow WW}$).
    \item Production of a scalar boson A decaying to a Higgs and a Z bosons. Higgs bosons are then forced to decay to ZZ. The final state is ZZZ ($A \rightarrow HZ \rightarrow ZZZ$).
\end{itemize}
The experimental setup employs a Cartesian coordinate system with a specific setup used for particle physics measurements. The z-axis is aligned in parallel to the beam direction, while the x and y axes together define the transverse plane of measurement. Key parameters are calculated as follows:
\begin{itemize}
    \item Azimuthal angle $\phi$ is determined with reference to the x-axis.
    \item Polar angle $\theta$ is measured from the positive z-axis.
    \item Pseudorapidity $\eta$ is computed using the expression: $\eta = \mathrm{log(tan(\theta/2)}$.
    \item The transverse momentum $p_T$ represents the momentum component projected onto the plane perpendicular to the beam axis.
\end{itemize}

For each event, we consider the two highest-$p_T$ jets in the event. Each jet is represented as a list of particle constituents, with $100$ highest-$p_T$ particles which are located inside the jet cone in $\Delta R = \sqrt{\Delta \phi^2 + \Delta \eta^2} < 0.8 $, where $\Delta \phi$ and $\Delta \eta$ are calculated from the jet axis. Each particle is represented as a vector of three features: $p_T$, $\Delta\eta$, and $\Delta\phi$. Full description of the dataset and its production can be found in Ref.~\cite{Belis2024}. As a result of this event processing, each event consists of two $100 \times 3$ feature matrices, which are then processed by the encoder described in Ref.~\cite{Belis2024} to derive the latent space representation of each jet.

The encoder of the autoencoder plays the role of a \textit{dimensionality reduction} method, as often used in various domains of data science. The model’s architecture consists of convolutional and fully-connected neural network layers, and it is described in detail in the work Ref.~\cite{Belis2024}. The autoencoder’s per-particle level training methodology produces an output dimension $l$ from the encoder that corresponds directly to the latent features of a single jet. The dataset passed further for analysis consists of dijet events, making the feature dimension $2l$.

\subsection*{Fundamental Concepts and Notation}
\subsubsection*{Tensor networks}
Tensor Networks are structured graphs composed of connected tensors - multilinear operators or simply multidimensional arrays of numerical values. It is convenient to represent TNs in a graphical representation where tensors are depicted as graph vertices, with indices representing connecting edges. The dimensionality of the tensor is defined through the number of indices or the tensor's rank. Connections between tensors symbolize contraction operations, which execute the summing of products across shared indices. The simplest example of contraction is matrix multiplication, which represents a specific case of order-2 tensor contraction. 

The computational strength of tensor networks lies in their ability to perform sophisticated, multi-layered mathematical operations. These include tensor contractions, inner product calculations, dimensional reduction with efficient low-rank representation, singular-value decomposition, eigenvalue decompositions, and outer product computations. Another important feature of a TN is its canonical form, which represents a normalization and standardization technique that provides a unique, reduced representation of that TN. Canonization enables more efficient computational analysis and comparison between TN configurations, minimizes redundancy among tensors, and reduces degrees of freedom. 

Various tensor network topologies have been explored in the field throughout the years. These include one-dimensional structures (Matrix Product States~\cite{MPS_cite}, Matrix Product Operators~\cite{MPO}), two-dimensional representations (Projected Entangled Pair States~\cite{PEPS}), and tree-like topologies (Tree TNs~\cite{Cirac_2021}). Each of these configurations offers different advantages for analyzing different types of data and physical systems.
This versatility makes TNs useful across multiple scientific domains, from quantum physics~\cite{MPS-PEPS-VRGM, ORUS2014117, Or_s_2019, tindall2025dynamicsdisorderedquantumsystems} to quantum computing~\cite{PhysRevA.101.032310, Synergistic, PhysRevResearch.6.013326} and ML~\cite{stoudenmire2017supervised, Reyes_2021, wang2025tensornetworksmeetneural, puljak2025tn4mltensornetworktraining}. They provide a powerful tool to manipulate complex computational problems by breaking them down into manageable mathematical representations.

\begin{figure}[htb]
    \centering
    \includegraphics[width=\linewidth]{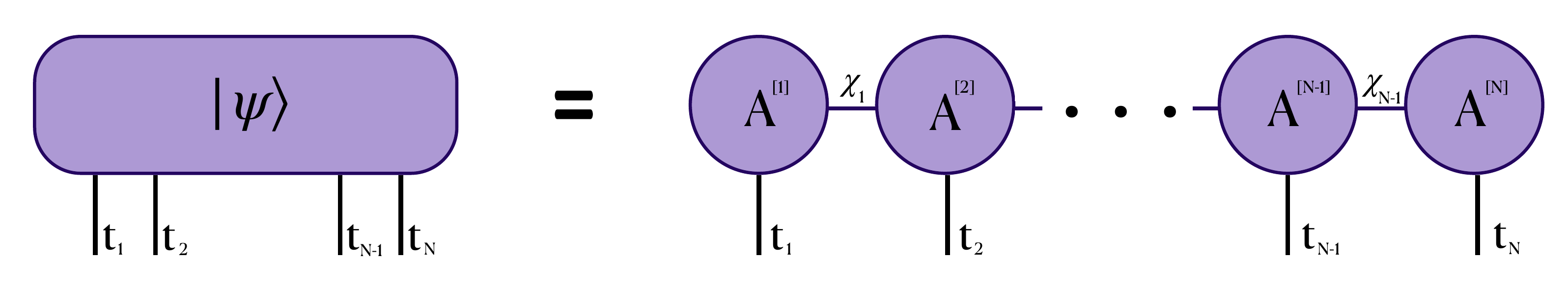}
    \caption{Graphical representation of a quantum state $|\Psi\rangle$ decomposed to an MPS with physical indices $t_k$ and bond indices $\chi_k$.}
    \label{fig:mps}
\end{figure}

\emph{Matrix Product State} (MPS), a one-dimensional TN model used in this study, is a low-rank representation of an N-order tensor or a quantum state $|\psi\rangle$. This structure is a chain or ring structure of 2-rank or 3-rank tensors, as visualized in Fig.~\ref{fig:mps}, where each tensor has a physical index with dimension $d$ and a connecting or bond index with dimension $\chi$. Using mathematical notation, we can represent the state $|\psi\rangle$ with an MPS decomposition:

\begin{equation}
    |\Psi\rangle = \sum_{\substack{t_1, t_2, \ldots, t_N \\ \chi_1, \chi_2, \dots, \chi_{N-1}}} A^{t_1}_{\chi_1} A^{t_2}_{\chi_1, \chi_2} \cdots A^{t_N}_{\chi_{N-1}}|t_1 t_2 \cdots t_N\rangle,
\end{equation}
where each matrix $A^{k}_{\chi_{k-1}, \chi_k}$ is a rank-3 tensor, and represents a coefficient of the $|t_k\rangle$ basis state. MPS with open boundary condition (obc) has the first and last tensor as rank-2. The space complexity of the N-order tensors is $\mathcal{O}(d^N)$, as opposed to the MPS complexity of $\mathcal{O}(Nd\chi^2)$, which proves its effectiveness.

\subsubsection*{Discrete Probabilistic Modeling with Tensor Networks}
In probabilistic anomaly detection, the goal is to learn the joint probability distribution of normal samples, such that low-probability inputs can be identified as anomalies. 
In the realm of quantum-inspired ML, \textit{Born Machines} represent a generative model that leverages the probabilistic nature of quantum mechanics to define probability distributions. This provides a powerful framework for modeling categorical or finite-state data. To capture this, we describe a discrete data setting where a quantum state $|\Psi\rangle$ represents an entire data distribution~\cite{aizpurua2024tensornetworksexplainablemachine}:
\begin{equation}
    |\Psi\rangle = \sum_{t \in \tau}{\Psi(t)|t\rangle}.
\end{equation}

In this formulation, $t$ represents a complete discrete data point (a tuple of $N$ random variables), $\tau$ is the set of all possible unique combinations of these $N$ random variables, $\Psi(t)$ represents the quantum amplitude associated with each unique data point configuration, and $|t\rangle$ denotes the basis states of the Hilbert space corresponding to each data point.

Using the definition of the Born rule, when we measure this quantum state, it collapses to a specific result $t \equiv (t_1, t_2, ..., t_N)$ with probability defined as:
\begin{equation} 
    P(t) = \frac{1}{Z}|\Psi(t)|^2, 
\end{equation}

Each $t_k$ corresponds to the observed value of the $k$-th discrete random variable in an $N$-dimensional Hilbert space. The normalization factor $Z$, also referred to as \textit{partition function}, ensures the probability distribution's validity:
\begin{equation} 
    Z = \sum_{t \in \tau} |\Psi(t)|^2.
\end{equation}

These equations ensure two fundamental properties of discrete probability distributions:
\begin{enumerate}
    \item Non-negative probabilities: $P(t) \geq 0$,
    \item Total probability normalization: $\sum_t P(t) = 1$.
\end{enumerate}

The coefficient $\Psi(t)$, which determines the probability of a given configuration, can be efficiently represented using an MPS~\cite{PhysRevX.8.031012, aizpurua2024tensornetworksexplainablemachine}. Instead of explicitly storing $\Psi(t)$ as a rank-$N$ tensor, we decompose it to an MPS, a chain structure of at most 3-rank tensors, using sequential singular value decomposition:

\begin{equation}
    \Psi(t_1, t_2, \ldots, t_N) = A^{(1)t_1} A^{(2)t_2} \cdots A^{(N)t_N},
\end{equation}

where each $A^{(k)t_k}$ is a matrix $\chi_{k-1} \times \chi_{k}$ that depends on the value of $t_k$ at position $k$. Boundary conditions are set with a vector $A^{(0)t_0}$ of dimension $d_0 \times \chi$ and $A^{(N)t_N}$ of dimension $\chi \times d_N$, where $d_k$ is a physical dimension of site $k$. The assumption is that the bond dimension $\chi$ between neighboring tensors is the same across all MPS sites.

Since the coefficient $\Psi(t)$ can be parametrized with MPS, the otherwise exponentially hard summation in $Z$ has the time complexity of $\mathcal{O}(Nd\chi^3)$, with $d=\mathrm{max}_id_i$~\cite{meiburg2024generativelearningcontinuousdata}. To reduce the complexity further, the TN can be put to a canonical form, ensuring $Z=1$ and $\mathcal{O}(N\chi^2)$. 

\begin{figure}[htb]
    \centering
    \includegraphics[width=\linewidth]{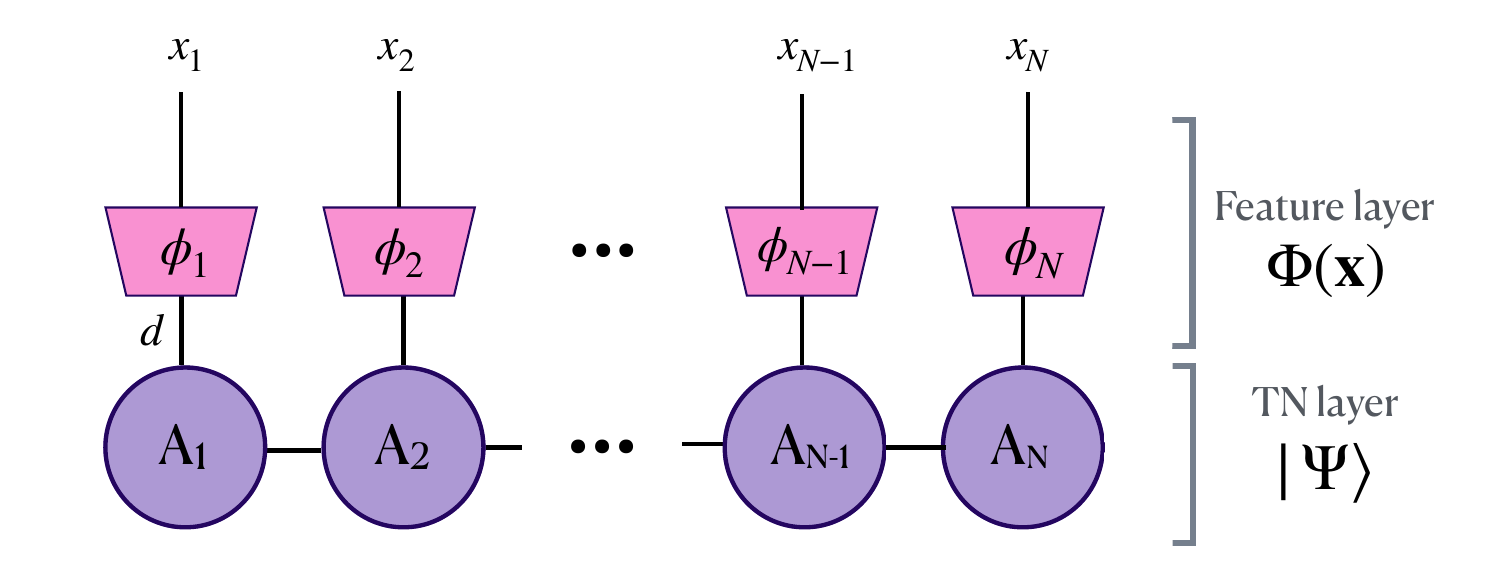}
    \caption{Graphical representation of the continuous-valued MPS model used for anomaly detection, consisting of a product state $\Phi(\mathbf{x})$ (feature layer), and a parametrized MPS $|\Psi\rangle$ with $A_k$ elements (TN layer).}
    \label{fig:mps_model}
\end{figure}

\subsection*{Tensor Network for Anomaly Detection in Continuous Data Regime}
To extend the probabilistic modeling to a continuous data setting, the Ref.~\cite{meiburg2024generativelearningcontinuousdata} proposes a continuous-valued MPS model (see Fig.~\ref{fig:mps_model}), consisting of:

\begin{enumerate}
    \item \textit{Feature layer}: a tensor product $\Phi$ of all feature maps $\phi_i(x_i)$ that maps each feature $x_i\in \mathbb{R}$ from a continuous domain to a $D$-dimensional discrete-valued vector space $\mathbb{K}^D$ ($\phi_i: \mathbb{R} \rightarrow \mathbb{K}^D$), where $i \in [1, N]$;
    \item \textit{Tensor Network}: a parametrized MPS $|\Psi\rangle$ that learns the probability distribution, with $N$ number of sites.
\end{enumerate}

Each \emph{feature map $\phi_i(x_i)$} is described with a set of feature functions $\mathcal{F} = \{f_j\}_{j=1}^D$ where each projection $f_j: \mathbb{R} \rightarrow \mathbb{K}$ creates an orthonormal basis for space $\mathbb{K}^D$. To ensure the continuous generalization of discrete probability modeling, $\phi_i$ mapping needs to be \textit{isometric}, which preserves distances between points in different spaces, as stated in Ref.~\cite{meiburg2024generativelearningcontinuousdata}. The choice of the appropriate feature map $\phi(x_i)$ depends on the feature domain, although it does not always lead to the best results. Some examples of isometric feature maps are the Fourier exponentials, Legendre, Laguerre, and Hermite polynomials. Ref.~\cite{meiburg2024generativelearningcontinuousdata} explains how, from an arbitrary feature function, one can create isometric maps. For a complete data point $\mathbf{x}$, the feature layer is defined as a product state:
\begin{equation}
    \Phi(\mathbf{x}) = \bigotimes_{i=1}^{N} \phi_i(x_i),
\end{equation}
where each map $\phi_i$ has a physical dimension $d$ that depends on $D$, and it can be equal or different per site $i$.

Finally, the continuous-valued MPS model, used to model arbitrary continuous-valued probability density function (PDF) $\theta(\mathbf{x})$, contains a product state $\Phi(\mathbf{x})$ contracted with a parameterized MPS $|\Psi\rangle$ with $N$ sites, physical dimension $d$, and bond dimension $\chi$ (see Fig.~\ref{fig:mps_model}). The PDF is described as:

\begin{equation}
    \theta(\mathbf{x}) = \langle \Phi(\mathbf{x})  | \Psi \rangle.
    \label{eq:theta}
\end{equation}

Following the Born rule formalism, the inner product in Eq.~\eqref{eq:theta} is transformed into the probability of an outcome $P(\mathbf{x})$ by taking the squared magnitude:
\begin{equation}
    P(\mathbf{x}) = |\langle \Phi(\mathbf{x})  | \Psi \rangle|^2,
    \label{eq:p_x}
\end{equation}

Non-negativity condition, $P(\mathbf{x}) \geq 0$, is satisfied directly from Eq.~\eqref{eq:p_x}. Moreover, $\int_{-\infty}^{\infty} P(\mathbf{x}) dx = 1$, is ensured by isometric conditions of $\Phi(\mathbf{x})$ and the requirement of the MPS having unit norm $\langle \Psi | \Psi \rangle = 1$. However, $\exists x \in \mathbb{R} : P(\mathbf{x}) > 1$ for particular points $x$.

\subsection*{Anomaly detection pipeline}
The first step of the anomaly detection pipeline consists of passing the jet training data through a trained convolutional autoencoder from~\cite{Belis2024} to obtain latent space features for a single jet with reduced dimensionality $l$. The last layer of the encoder model has \textit{tanh} activation function, thus features are on the domain range $[-1, 1]$. For further analysis, we use the dijet dataset, consisting of $2l$ features, which we denote as $x$. These features are embedded into a Product State $\Phi(x)$ using the Laguerre polynomial feature map with degree $n$: 
\begin{equation}
    L_n(x) = \sum_{k=0}^{n} \frac{(-1)^k}{k!} \binom{n}{k} x^k,
\end{equation}
following the procedure for embedding described in Ref.~\cite{puljak2025tn4mltensornetworktraining}. Here, the embedding directly defines the physical dimension $d$ of the TN model, depending on the degree $n$, such that each feature $x_i$ is embedded with $L_n(x_i)$ resulting in a product state:
\begin{equation}
    \Phi(x) = \bigotimes_{i=1}^{N} L_n(x_i).
\end{equation}

Next, we define the parametrized MPS model $|\Psi\rangle$ and train it by optimizing the parameters of $A^{(k)^{t_k}}$ with Mini-Batch Gradient Descent to maximize the likelihood of the training data. Specifically, we use the gradient descent implementation from the \texttt{tn4ml}~\cite{puljak2025tn4mltensornetworktraining} library to minimize the Negative Log-Likelihood (NLL) function:
\begin{equation}
    L = -\frac{1}{|D|} \sum_{x \in D} \ln P(x),
    \label{eq:NLL}
\end{equation}
where $D$ is the dijet dataset and $P(x) = |\langle \Phi(x)  | \Psi \rangle|^2$ is the probability of given input sample. $|\Psi\rangle$ and $|\Phi(x)\rangle$ are normalized to keep numerical stability during the training.

Certain feature values may cause deviations from the model's learned representation of normal behavior. In such cases, a low assigned probability serves as an indication of a potential anomaly. In contrast, feature values with high probabilities align with the model's expected patterns, reflecting normal behavior.
This probabilistic framework can be used to rank events according to their likelihood under the model describing the dataset (standard events), allowing low-probability instances to be selected as candidates for a dataset enriched in potentially anomalous events.
Furthermore, this method provides a direct anomaly detection metric that can be used in real-time applications. The small inference time allows for instantaneous anomaly detection without requiring complex statistical analyses.

\section{Results and Analysis}\label{sec:results}
The implementation of the full training and evaluation pipeline is implemented using the library \texttt{tn4ml}~\cite{puljak2025tn4mltensornetworktraining}, which facilitates the development of TNs as ML models. Optimization is performed using Mini-batch Gradient Descent with Adam optimizer, $N_{\mathrm{train}} = 10^5$ and learning rate $10^{-4}$. The performance is evaluated in three stages before choosing the final setup of hyperparameters.

The autoencoder from Ref.~\cite{meiburg2024generativelearningcontinuousdata} can be used alone for the anomaly detection task, achieving $75\%$, which serves as our baseline. We investigate whether one can improve this result using the TN-based model in the pipeline.

\emph{\textbf{Stage 1 }} We benchmark the anomaly detection capabilities against the $A \rightarrow HZ \rightarrow ZZZ$ model, and evaluate the model's performance at latent dimensions $l = 4$, where the number of input features is $2l$.
Since the final activation function of the encoder part of the AE used for dimensionality reduction is \textit{tanh}, the latent features are continuous and constrained to the range $[-1, \ 1]$. Following the intuition from Ref.~\cite{meiburg2024generativelearningcontinuousdata}, which suggests using Legendre polynomials for data defined on this range, we train the model using different initializers together with this specific embedding to analyze the variability of the training optimization. 

\begin{figure}[htb]
    \centering
    \includegraphics[width=\linewidth]{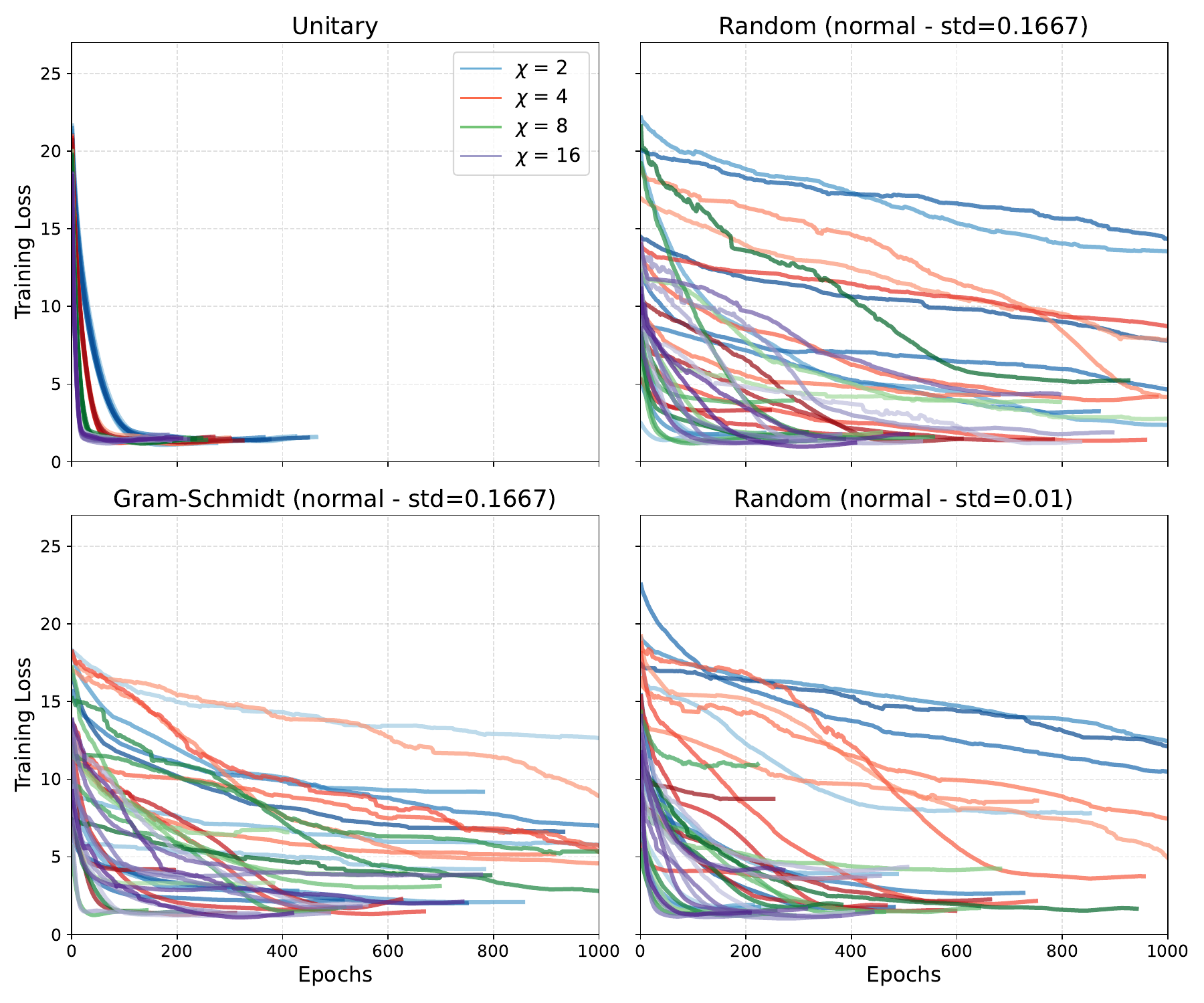}
    \caption{Negative Log Likelihood loss as a function of the training epoch for each initializer and bond dimension $\chi$ using a Legendre feature map of degree two.}
    \label{fig:train_loss}
\end{figure}

For the number of tensors $L=2l$, we assess performance for bond dimensions $\chi \in [2, 4, 8, 16]$. 
Fig.~\ref{fig:train_loss} shows the training NLL loss for the following hyperparameters: (1) bond dimension $\chi \in [2, 4, 8, 16]$; (2) initializers - unitary; random normal with standard deviation $\sigma=(\chi_{\mathrm{max}} \: \cdot \: d)^{-1}$; random normal with $\sigma=10^{-2}$; and Gram-Schmidt orthogonalization with normal distribution $\sigma=(\chi_{\mathrm{max}} \: \cdot \: d)^{-1}$, where $d$ is the embedding dimension. Each configuration is trained ten times to ensure statistical robustness and to account for variation in model performance due to random initialization and shuffling of the data samples. The color palette in subplots in Fig.~\ref{fig:train_loss} corresponds to the bond dimension, with five distinct shades representing individual runs. The maximum number of training epochs is 1000 with early stopping and patience of 30 epochs. From Fig.~\ref{fig:train_loss}, we observe significant variability in training behavior for some initializers. In particular, initializers with higher standard deviation result in more unstable and slower convergence. Given these findings, we select the most stable option, the \textit{unitary initializer}, which initializes tensors as stacks of random unitary matrices.

\begin{figure*}[htb]
    \centering
    \includegraphics[width=\linewidth]{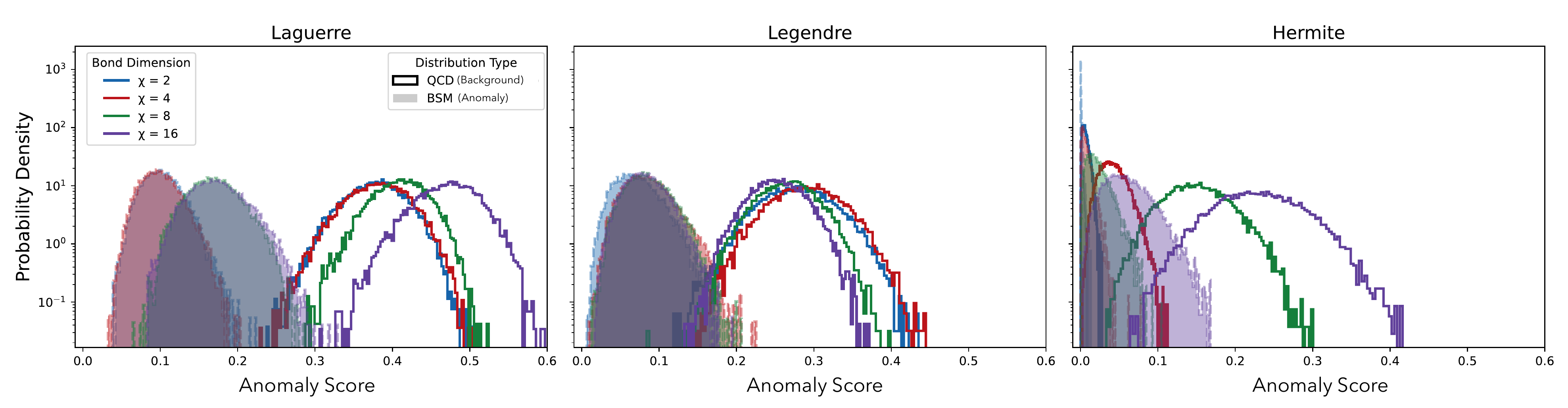}
    \caption{Comparison of anomaly scores for QCD (background) and BSM (anomaly) across different embedding polynomial functions - Laguerre (left), Legendre (center), and Hermite (right) for latent space $l=4$. Each distribution corresponds to a different bond dimension $\chi \in {2, 4, 8, 16}$. The solid line style visualizes QCD distribution, and the filled histogram is for BSM.}
    \label{fig:ad_scores}
\end{figure*}

\emph{\textbf{Stage 2 }} To ensure the embedding is appropriately chosen, we also evaluate the performance using two additional feature maps: Hermite polynomials, defined over the entire $\mathbb{R}$ space, and Laguerre polynomials, defined on the non-negative real axis. For the Laguerre map, the data is rescaled to the range $[0, 1]$. All three feature maps are used with polynomial expansion of degree two. To asses anomaly detection performance, we compare the distributions of anomaly scores(i.e., the output probabilities of SM and BSM events) in Fig.~\ref{fig:ad_scores} for different bond dimensions $\chi$ and embedding methods. From these plots, we conclude that the best separation between signal and background events is achieved using Laguerre polynomials. 

\emph{\textbf{Stage 3 }} Based on these observations, we study the performance of the MPS model by computing the Receiver Operating Characteristic (ROC) curve with the corresponding Area Under the Curve (AUC) value. To gain deeper insights, we further examine the key metrics commonly used in HEP analysis: the true positive rate (TPR), or signal efficiency ($\varepsilon_s$), and the corresponding false positive rate (FPR), or background efficiency ($\varepsilon_b$). In particular, we focus on the inverse background efficiency $\varepsilon_b^{-1}$ at specific signal efficiency values $\varepsilon_s \in {0.6, 0.8}$, following the notation from Ref.~\cite{Belis2024}. 

\begin{figure}[htbp]
    \centering
    \includegraphics[width=0.8\linewidth]{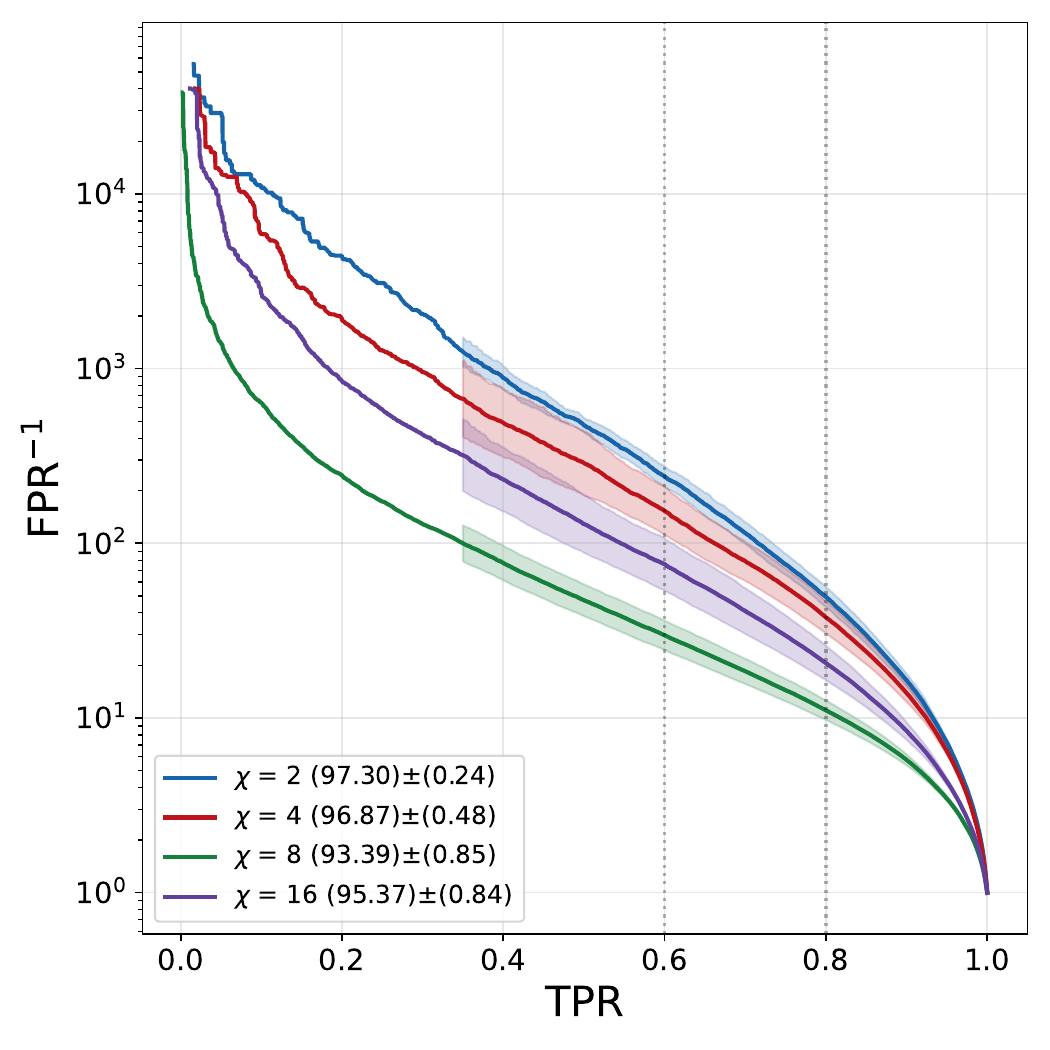}
    \caption{ROC curve for MPS model for latent space $l=4$ with unitary initializer, Laguerre polynomial feature map and bond dimensions $\chi \in [2, 4, 8, 16]$. Error bands around each ROC curve represent the RMS spread of the results induced by repeating the training and testing runs with different random initialization and seeds for data shuffling. Smaller test statistics in area of low TPR values result in large error bands, which are discarded from the plot.}
    \label{fig:roc_1}
\end{figure}

Fig.~\ref{fig:roc_1} provides insights into the optimal hyperparameter configuration, highlighting that a bond dimension $\chi = 2$ achieves the best overall results. While higher bond dimensions in theory offer greater expressive power, our findings indicate that this is not valid for models with a low number of tensors, as they suffer from overfitting to the training data and fail to generalize effectively when evaluated on the test dataset. This demonstrates the advantage of using a simpler model, with fewer parameters, which may also be suitable for on-chip deployment. In contrast, for larger latent space dimensions ($l=8$ or $16$), increasing the bond dimension initially improves performance, up to a specific \textit{sweet spot} beyond which the performance begins to decrease. For both latent space dimensions, this optimal point occurs at a bond dimension of $\chi = 32$.
Additionally, we studied the impact of increasing the polynomial degree in the feature map, with the conclusion that the higher degree does not bring better results, setting it to $\mathrm{degree}=2$.

\emph{\textbf{Final Setup }} We test the best model’s setup for anomaly detection across data-dependent parameters: latent space dimensions $l$, and different new physics scenarios to determine whether the model exhibits any bias toward specific signals or if its conclusions remain consistent across new-physics scenarios. 

\begin{figure}[htbp]
    \centering
    \includegraphics[width=0.8\linewidth]{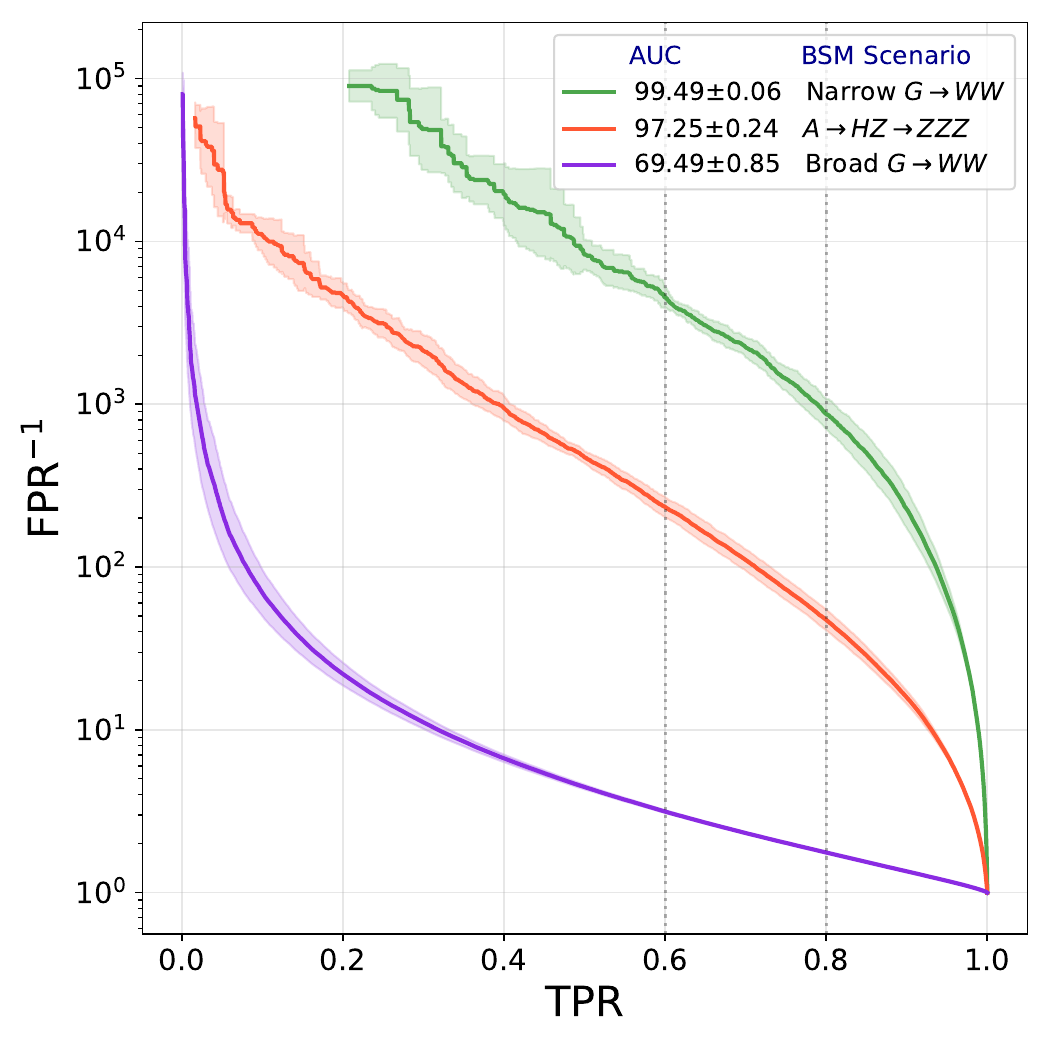}
    \caption{ROC curve for MPS model for latent space $l=4$ with unitary initializer, Laguerre polynomial feature map and bond dimension $\chi = 2$ for different BSM scenarios. Error bands represent the RMS spread of the results induced by repeating the training/testing runs with different random seeds for initialization and data shuffling. Smaller test statistics in area of low TPR values result in large error bands, which are discarded from the plot. }
    \label{fig:roc_signals}
\end{figure}

Fig.~\ref{fig:roc_signals} visualizes the performance for the three BSM scenarios mentioned above. Different performance across different signals is achieved due to the nature of the new-physics process, with the broad Graviton being the hardest one to distinguish from the background distribution. 
Our results for the latent space $l=4$ are comparable to the results of the quantum unsupervised kernel method (QKM) for the latent space $l = 8$ in Ref.~\cite{Belis2024}, where the quantum circuit for kernel estimation has number of qubits $n_q = l$ and three repetitions of data re-uploading scheme with near-neighbor entanglement. This shows that even with the latent space $l=4$ we can achieve competitive results for anomaly detection using smaller dimensionality of the input data, and a less complex model. The most interesting finding are the results for the signal $\mathrm{BR \ G \rightarrow WW}$, with AUC of $69.49\pm0.85\%$, which significantly outperforms the QKM ($47.62\pm0.52\%$ AUC), with an absolute gain of $21.87$ percentage points in AUC.

\begin{figure}[htbp]
    \centering
    \includegraphics[width=0.8\linewidth]{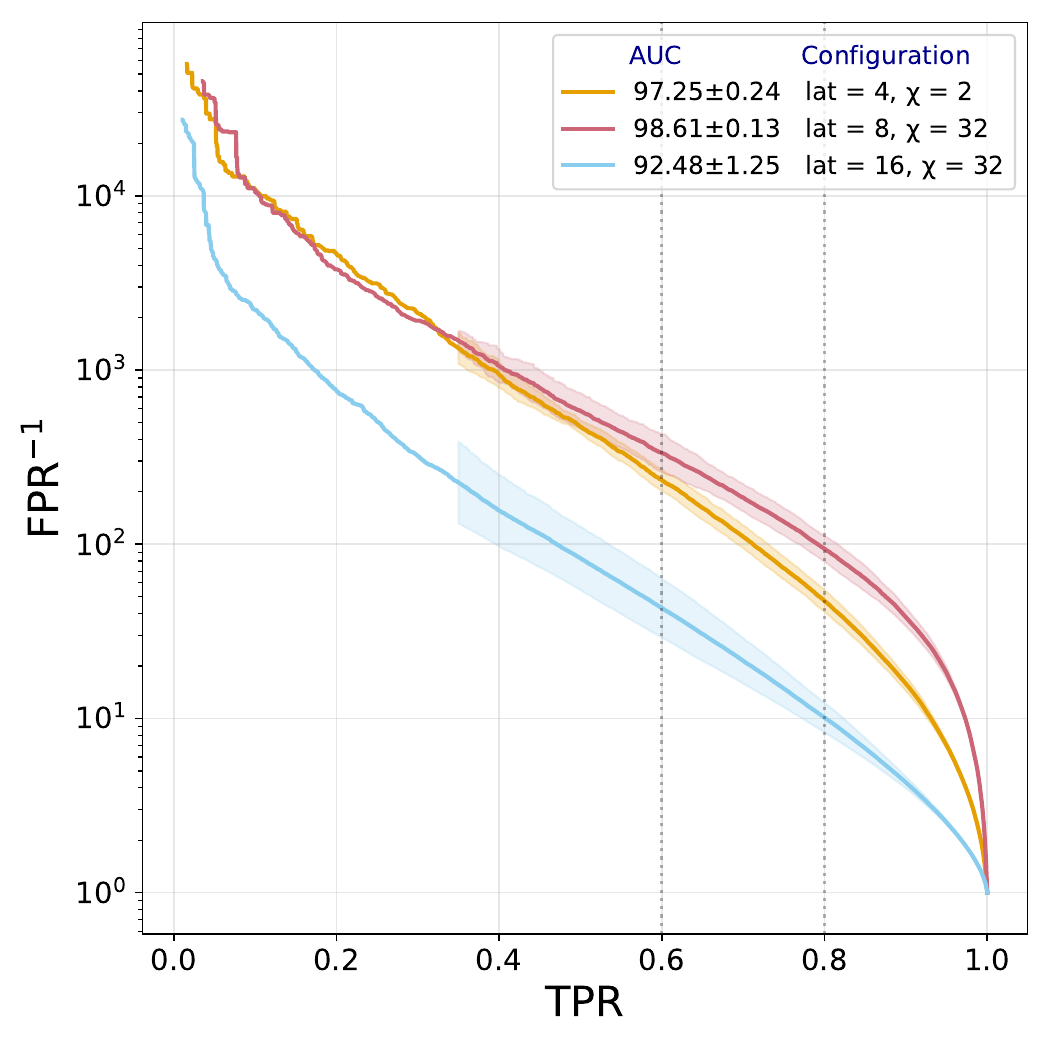}
    \caption{ROC curve for MPS model with unitary initializer, Laguerre polynomial feature map for different latent space dimension $l \in [4, 8, 16]$. Error bands represent the RMS spread of the result induced by repeating training/testing runs with different random initialization and seeds for data shuffling. Smaller test statistics in areas of low TPR values result in large error bands, which are discarded from the plot.}
    \label{fig:roc_2}
\end{figure}

Fig.~\ref{fig:roc_2} shows the performance of the model on different latent spaces, for signal $A\rightarrow HZ \rightarrow ZZZ$. For each latent space, we choose the best performing bond dimension, e.g., for $l=4 \rightarrow \chi = 2$, $l=8 \rightarrow \chi = 32$, and $l=16 \rightarrow \chi = 32$. There exists a saturation of performance in the model size controlled by the latent space dimension $l$ and bond dimension $\chi$. The highest performance is achieved for latent space $l=8$ and bond dimension $\chi=32$. To align with the findings of Ref.~\cite{Belis2024}, Table~\ref{tab:efficiency} shows the values of $\varepsilon_{b}^{-1} \text{ at signal efficiencies } \varepsilon_s \in \{0.6, 0.8\}$ for latent space dimensions $l \in {4, 8, 16}$. These results numerically support the trends observed in Figs.~\ref{fig:roc_signals} and \ref{fig:roc_2} for anomalous signature $A\rightarrow HZ \rightarrow  ZZZ$, contributing to the conclusion that performance saturates beyond a specific model complexity. Additionally, the table includes results for the anomalous process $\mathrm{BR \ G \rightarrow WW}$, which exhibits saturation already at a lower latent space dimension $l=4$. This observation further confirms the suitability of less complex architectures for achieving high sensitivity, enabling effective detection even of highly anomalous signatures.

\begin{table}[htbp]
\centering
\begin{tabular}{llcc}
\multicolumn{4}{c}{$A\rightarrow HZ \rightarrow ZZZ$} \\
\toprule
$l$ & $\chi$ & $\varepsilon_b^{-1}(\varepsilon_s=0.8)$  & \: $\varepsilon_b^{-1}(\varepsilon_s=0.6)$ \\
\midrule
    4 & 2 & 47$\pm$8 & 233$\pm$33\\
    8 & 32 & 94$\pm$17 & 334$\pm$85\\
    16 & 32 & 10$\pm$3 & 44$\pm$18 \\
\bottomrule
\\
\multicolumn{4}{c}{$\mathrm{BR \ G \rightarrow WW}$} \\
\toprule
$l$ & $\chi$ & $\varepsilon_b^{-1}(\varepsilon_s=0.8)$  & \: $\varepsilon_b^{-1}(\varepsilon_s=0.6)$  \\
\midrule
    4 & 2 &1.76$\pm$0.02 & 3.15$\pm$0.12\\
    8 & 32 &1.55$\pm$0.06 & 2.32$\pm$0.15\\
    16 & 32 & 1.39$\pm$0.04 & 1.92$\pm$0.12\\
\bottomrule
\end{tabular}
\caption{Comparison of performance metrics $\varepsilon_b^{-1}$ at $\varepsilon_s \in \{0.6, 0.8\}$ across latent spaces and signal types. $\chi$ is the corresponding bond dimension per model. Uncertainties are equal to standard deviations over five training and testing runs with different random initializations and data shuffling.}
\label{tab:efficiency}
\end{table}

\begin{figure}[ht]
    \centering
    \includegraphics[width=0.8\linewidth]{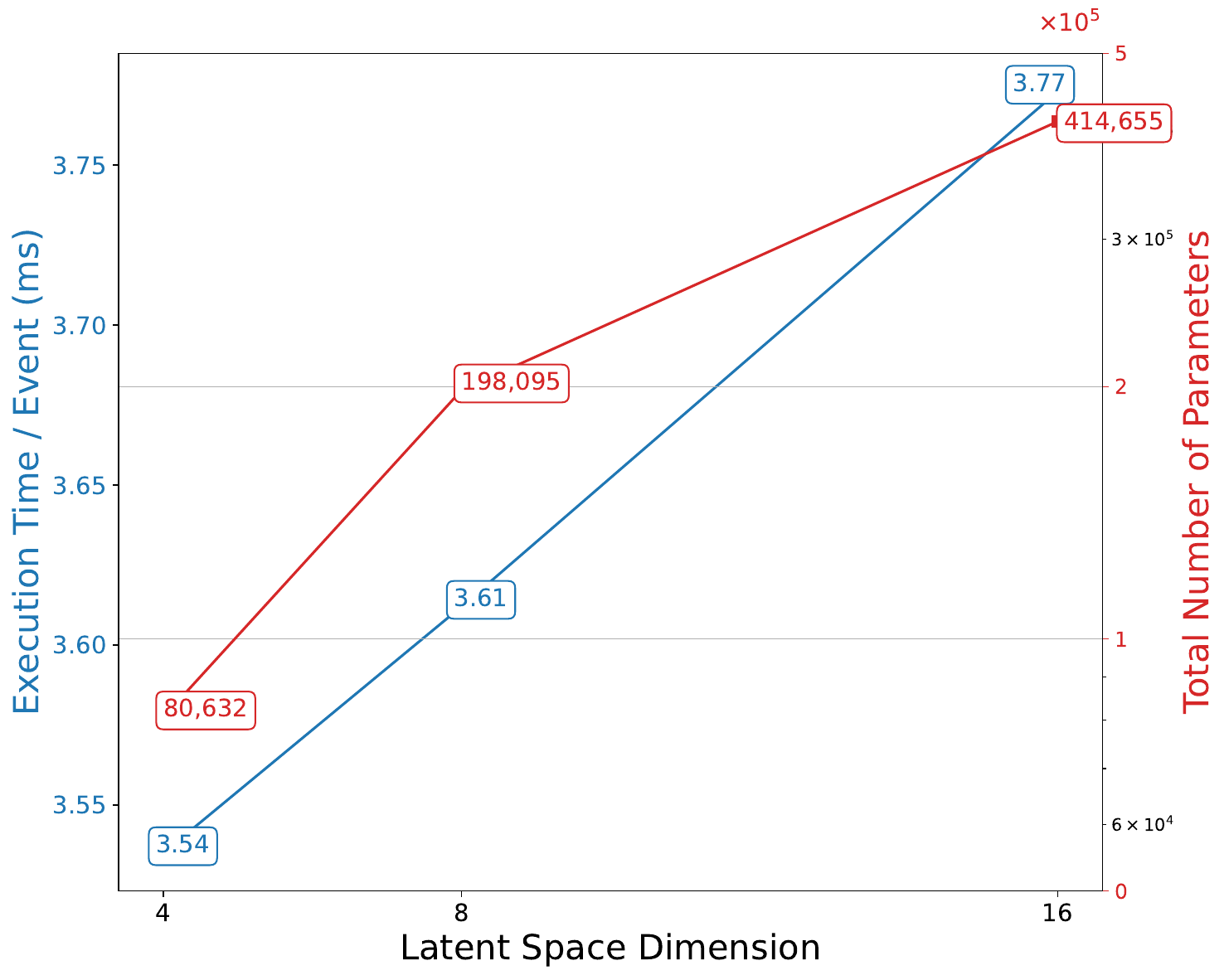}
    \caption{Execution time per batch size (ms) and number of parameters of the autoencoder and the Matrix Product State as a function of the latent space dimension. With the increase in model complexity (latent space), the inference time of the full pipeline rises to $3.77$ms.}
    \label{fig:runtime}
\end{figure}

\newpage
\emph{\textbf{Runtime Evaluation }} We measured per-event inference time for the full pipeline, containing both the encoder and the MPS model. We benchmark the execution in microseconds (ms) per event and model size, for three model configurations from Fig.~\ref{fig:roc_2} on an Intel Core i5-9600KF using one thread. All models, including the largest model with latent space $l=16$, are runnable in software under $\sim 100\text{ms}$ (see Fig.~\ref{fig:runtime}). This makes them suitable for deployment within the software-based HLT selection system. It is important to note that the HLT computing farm utilizes Intel Xeon CPUs, with higher memory bandwidth, larger caches and more parallel throughput, which could potentially further reduce the latency. 

\section{Conclusions}\label{sec:conclusion}
In this study, we presented a novel application of quantum-inspired Tensor Networks for anomaly detection in the latent space of proton collision events at the LHC. By leveraging the compressed latent space representations generated by an autoencoder and constructing a continuous-valued Matrix Product State model, we developed a framework capable of learning the probability distribution of Standard Model events and detecting deviations that indicate the presence of new-physics signatures. Following and building on the previous work with quantum kernel methods, we demonstrated that a TN model, when properly initialized and using an appropriate isometric feature map, can achieve comparable or improved anomaly detection performance even with smaller latent space dimensionality. Our systematic hyperparameter optimization of the MPS model, including feature map embeddings, initialization functions, and bond dimensions, showed that the Laguerre polynomial embedding with a unitary initializer provided the most stable training and best signal-background separation. Furthermore, this analysis demonstrates that lower-dimensional settings leverage low bond dimensions, which minimizes model complexity while preserving excellent sensitivity for anomalous physics scenarios. Importantly, the model shows the robustness and adaptability to different types of Beyond Standard Model anomalies. Our findings on inference times in a software-based environment indicate that the full pipeline can operate efficiently and is compatible with the latency requirements of the HLT system. To address the stricter latency and resource constraints of the FPGAs in the L1T, while balancing model performance, reducing latency through methods such as pruning and quantization remains a key objective.

This work presents TNs as robust ML models and positions them as a practical and powerful tool for enhancing anomaly detection performance in the high-energy physics experiments at the LHC. Future directions include extending this study for more complex new-physics signatures, porting the TN model onto FPGA architecture for real-time selection, and exploring further possibilities for enhancing practical QML methods.

\section*{Code Availability}
%\vspace{-1.2em}
The code is open-source and publicly available at\\\textcolor{blue}{\href{https://github.com/bsc-quantic/tn4ml/tree/6ee6f5f4dc525887014cd81025016f27aa156e15/docs/source/examples/tnad_latent}{github.com/bsc-quantic/tn4ml/docs/examples/tnad\_latent}}.
%\vspace{-1.2em}
\section*{Acknowledgements}
%\vspace{-1.2em}
We thank Vasilis Belis and Michele Grossi for useful discussions. E.P. was supported by CERN through the Quantum Technology Initiative in earlier stages of research. A.G.S. acknowledges financial support from the Spanish Ministry for Digital Transformation and of Civil Service of the Spanish Government through the QUANTUM ENIA project call - Quantum Spain, EU through the Recovery, Transformation and Resilience Plan – NextGenerationEU within the framework of the Digital Spain 2026.

\bibliography{bib}

\end{document}